%% file: CP_38_main.tex
\documentclass[runningheads]{llncs}
\usepackage{cite}
\usepackage{calrsfs}
\usepackage{graphicx}
\usepackage{tabularx}
\usepackage{xcolor}
\usepackage{subcaption}
\usepackage{booktabs}
\usepackage[english]{babel}
\usepackage{float}
\usepackage{framed}
\usepackage{algorithm}
\usepackage{algorithmic}
\usepackage{lipsum}
\usepackage{hyperref}
\newcommand*{\rd}{}
\newcommand{\shisha}{\texttt{Shisha}}
\begin{document}
\title{\shisha\/: Online scheduling of CNN pipelines on heterogeneous architectures}
%
%

\author{Pirah Noor Soomro\inst{1} \and Mustafa Abduljabbar\inst{2} \and \\ Jeronimo Castrillon\inst{3} \and Miquel Pericàs\inst{1}} 
\institute{Dept Computer Science and Engineering, Chalmers University of Technology  
\email{\{pirah,miquelp\}@chalmers.se}
\and Ohio State University 
\email{abduljabbar.1@osu.edu}
\and Chair for Compiler Construction, Technische Universit\"at Dresden \email{jeronimo.castrillon@tu-dresden.de}}
\authorrunning{P. Noor Soomro et al.}
%
\maketitle              
\vspace{-0.8cm}
\begin{abstract}

\sloppy 
Many modern multicore processors integrate asymmetric core clusters.
With the trend towards Multi-Chip-Modules (MCMs) and interposer-based packaging technologies, platforms will feature heterogeneity at the level of cores, memory subsystem and the interconnect.
Due to their potential high memory throughput and energy efficient core modules, these platforms are prominent targets for emerging machine learning applications, such as Convolutional Neural Networks (CNNs). 
To exploit and adapt to the diversity of modern heterogeneous chips, CNNs need to be quickly optimized in terms of scheduling and workload distribution among computing resources. To address this we propose \shisha{}, an online approach to generate and schedule parallel CNN pipelines on heterogeneous MCM-based architectures. \shisha\ targets heterogeneity in compute performance and memory bandwidth and tunes the pipeline schedule through a fast online exploration technique. 
We compare \shisha\ with Simulated Annealing, Hill Climbing and Pipe-Search.
On average, the convergence time is improved by $\sim35\times$ in \shisha\ compared to other exploration algorithms.
Despite the quick exploration, \shisha{}'s solution is often better than that of other heuristic exploration algorithms.

\end{abstract}

\input{CP38_introduction}
\input{CP38_motivation}
\input{CP38_background}
\input{CP38_relatedWork}
\input{CP38_methodology}
\input{CP38_experimental_setup}
\input{CP38_evaluation}

\input{CP38_conclusion}
\bibliographystyle{splncs04}
\vspace{-.5cm}
\bibliography{CP38_references.bib}
%
%
%

%
%
%

%
\end{document}

%% file: CP38_introduction.tex
\vspace{-.8cm}
\section{Introduction}\label{sec:intro}\vspace{-.3cm}


Multicore processors are becoming more and more heterogeneous. 
Intel's Meteor Lake~\cite{intel_2022} features asymmetric multicore design containing
high performance and power saving cores. Similarly, Apple's A14 Bionic~\cite{nanoreview.net} integrates high performance cores called Firestorm and power saving cores called Icestorm. 
The trend towards heterogeneity is complemented with the trend towards Multi-Chip-Module (MCM) integration, which enables lower cost during design and improves yield by reducing chip area (chiplets)~\cite{kannan2015enabling}. 
When combined with interposer-based packaging technology, it enables lower latency and high bandwidth transmission to memory devices such as High Bandwidth Memory (HBM)~\cite{7383697}.
Chip manufacturers are adopting a mix of these technologies in order to design high performance processors, resulting in heterogeneity at the level of the cores, memory subsystem and the Network on Chip (NoC).
In order to effectively exploit such architectures, applications must be optimized considering the impact of different levels of heterogeneity. Furthermore, to address the diversity of hardware platforms, the optimization process must be fast and preferably online.  
Convolutional Neural Networks (CNNs) have high computational, bandwidth and memory capacity requirements owing to the large amount of weights
 and the increasing size of intermediate results that need to be transferred between layers. 
Parallel pipelining has the potential to address these requirements by partitioning the whole network across devices, and requiring only the inputs to be exchanged among stages. In chiplet architectures, CNNs could be efficiently pipelined by distributing layers across chiplets so as to reduce the amount of weights that need to be copied. Furthermore, pipelining makes the task of load balancing manageable among heterogeneous computing units.

In order to partition and schedule pipelines, current approaches rely on
designing cost models to steer design space exploration~\cite{adams2019learning,anderson2021efficient}.
For instance, the auto-scheduler in~\cite{adams2019learning} explores over ten thousand schedules for a single CNN-layer pipeline using Halide~\cite{ragan2013halide}.
The effectiveness of these approaches depends on the accuracy of the cost model and the scalability of the exploration algorithm.
Sophisticated cost models, some of them using ML-models themselves, have been proposed and used in~\cite{mullapudi2016automatically, adams2019learning, li2021chimera, ahn2020chameleon, zheng2020ansor, wu2020pipeline, pipeit}.
These models, however, require extensive training for near-optimal solutions~\cite{anderson2021efficient}, are sensitive to changes in the execution environment (e.g., DVFS) and architectural parameters, need in-depth architectural knowledge for model updates, and do not consider the impact of heterogeneous multicore or chiplet architectures. 
As heterogeneity at different levels of processing (e.g. core performance, memory bandwidth and/or MCM organization) is expected to increase in future HPC platforms, static pipeline partitioning and scheduling become inflexible.
Online auto-tuning of the pipeline schedule would help to ensure performance portability to future architectures. However, to make it practical, it is critical that online pipeline partitioning and scheduling finds an acceptable configuration with low overhead.
Pipe-Search~\cite{soomro2021online} adopts an online exploration approach for finding a pipeline configuration. It generates a database of pipeline configurations which is space-intensive and prohibitively slow for larger systems and deeper CNNs.
In this paper, we propose a quick method to determine a meaningful starting point, or seed, for the exploration coupled with a simple navigation heuristic for efficient runtime auto-tuning.
In \shisha{}, we leverage statically available information from the CNN and from the target platform to reduce the number of exploration points and find a near-optimal solution within reasonable time.
A \emph{configuration} explored by \shisha\ suggests grouping CNN layers into pipeline stages and mapping of pipeline stages onto available sets of processing units referred to as \textit{Execution Places (EPs)}.
When generating initial configurations, \shisha\ aims at balancing the load among pipeline stages while considering the allocation of stages to EPs.
\shisha\ improves upon related work in two ways:
\begin{itemize}
    \item \shisha{} achieves faster convergence by introducing two novel schemes: (i) the seed generation and (ii) the online tuning. We demonstrate that \shisha\ is able to converge faster than existing algorithms (Simulated Annealing, Hill Climbing and Pipe-Search) and that it is able to find a solution within practical time limits.
    \item We show that \shisha\ scales better with deeper CNNs and 
    with larger amount of EPs per processing unit which is one of the limitations of prior online tuning approaches such as Pipe-Search~\cite{soomro2021online}.
\end{itemize}
\shisha\ maps pipeline stages to EPs, which 
could be of any type  and number of processing units, such as multicores or manycores. 
To measure the quality of schedules explored by \shisha\ we compare our results to conventional search exploration algorithms such as Simulated Annealing (also used by TVM~\cite{zheng2020ansor}), Hill Climbing, Exhaustive Search and Random Walk (executed for a longer period of time), and to Pipe-search, an earlier online tuning approach.
We test \shisha\ on state of the art CNNs such as ResNet50~\cite{he2016deep} and YOLOv3~\cite{redmon2018yolov3}. 
The results show that, despite exploring only a tiny portion of the design space ($\sim0.1\%$ of design space for ResNet50 and YOLOv3), 
\shisha\/ finds a solution that is equivalent to exhaustive search.
Moreover, due to the guided exploration, the convergence time is improved by $\sim35\times$ in \shisha\ compared to the other representative exploration algorithms.

%% file: CP38_motivation.tex
\vspace{-.5cm}
\section{Motivation and problem  definition}\label{sec:motivation}
\begin{figure}[!tb]
\centering
\begin{subfigure}[]{.49\linewidth}
\includegraphics[width=\linewidth,]{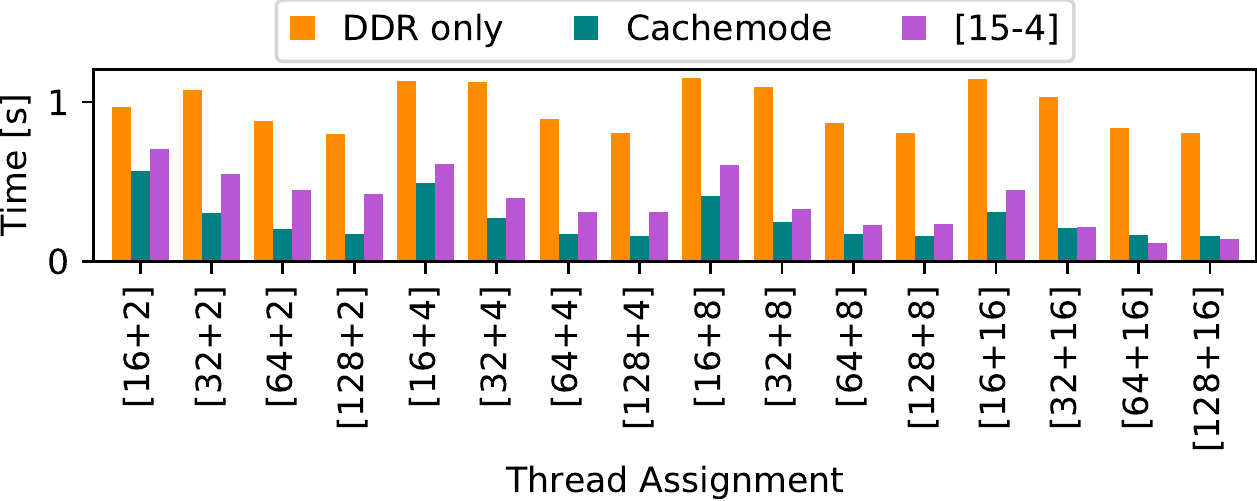}
\caption{[19GB]}\label{fig:19GB_cmp}
\end{subfigure}
\begin{subfigure}[]{.49\linewidth}
\includegraphics[width=\linewidth]{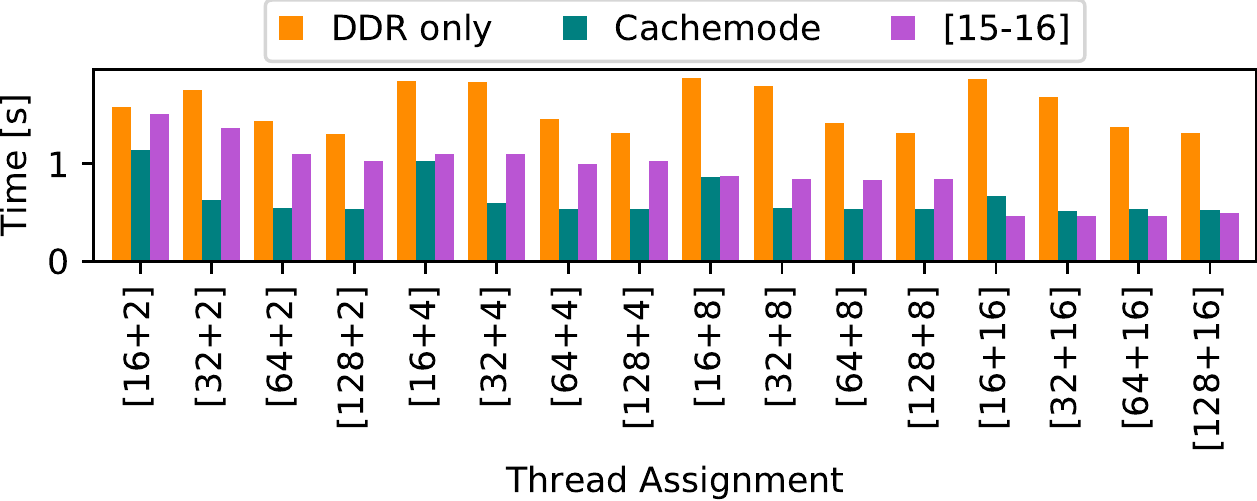}
\caption{[31GB]}\label{fig:31GB_cmp}
\end{subfigure}
\caption{Comparison of cases 1,2 and 3. X-axis [X+Y] shows X = threads assigned to MCDRAM and Y = threads assigned to DRAM}
\label{fig:comparisons}
\end{figure}

\begin{figure}
\begin{minipage}[]{0.48\linewidth}
\begin{subfigure}[]{.48\linewidth}
\includegraphics[width=\linewidth]{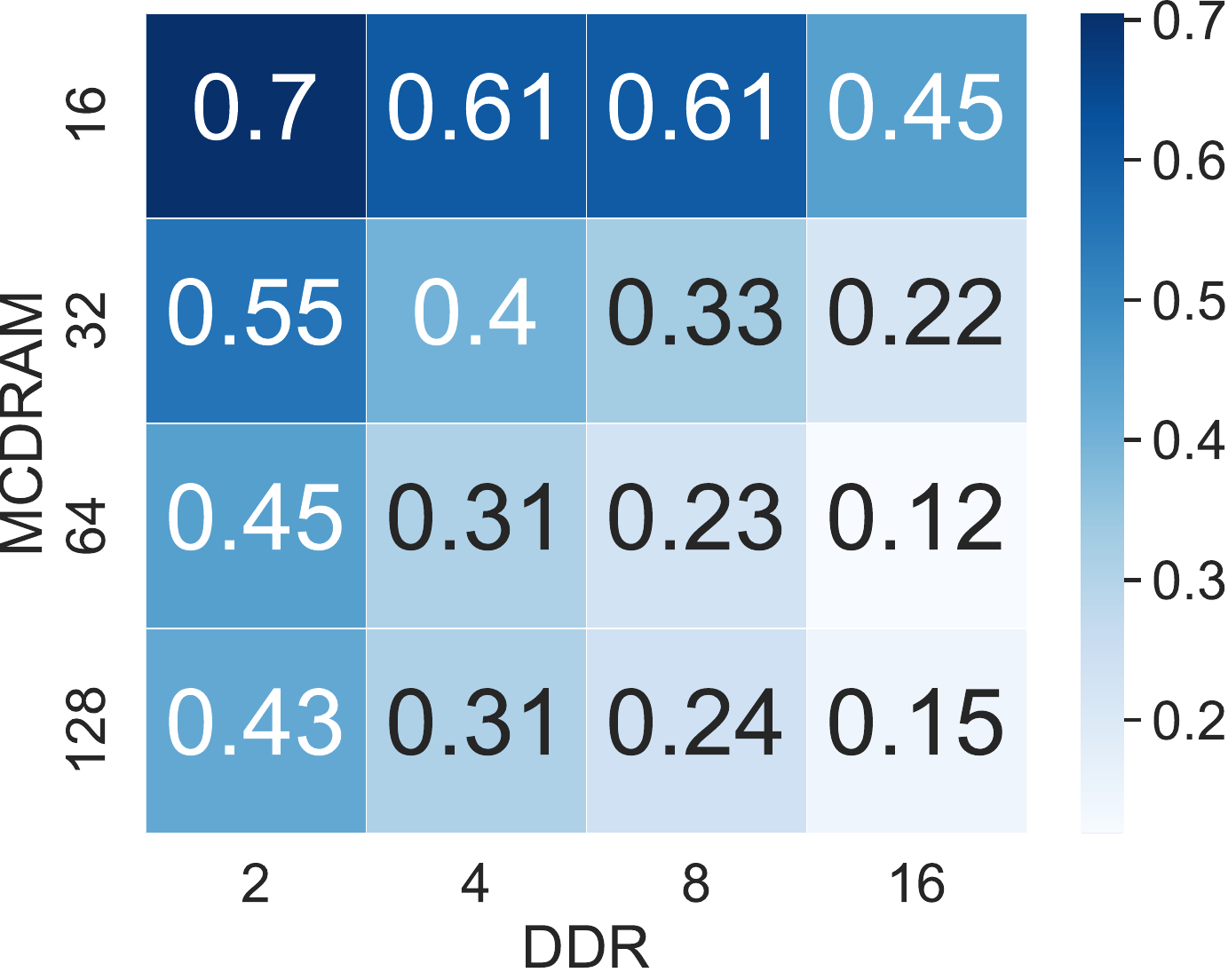}
\caption{[15-4]}\label{fig:19GB_time}
\end{subfigure}
\begin{subfigure}[]{.48\linewidth}
\includegraphics[width=\linewidth]{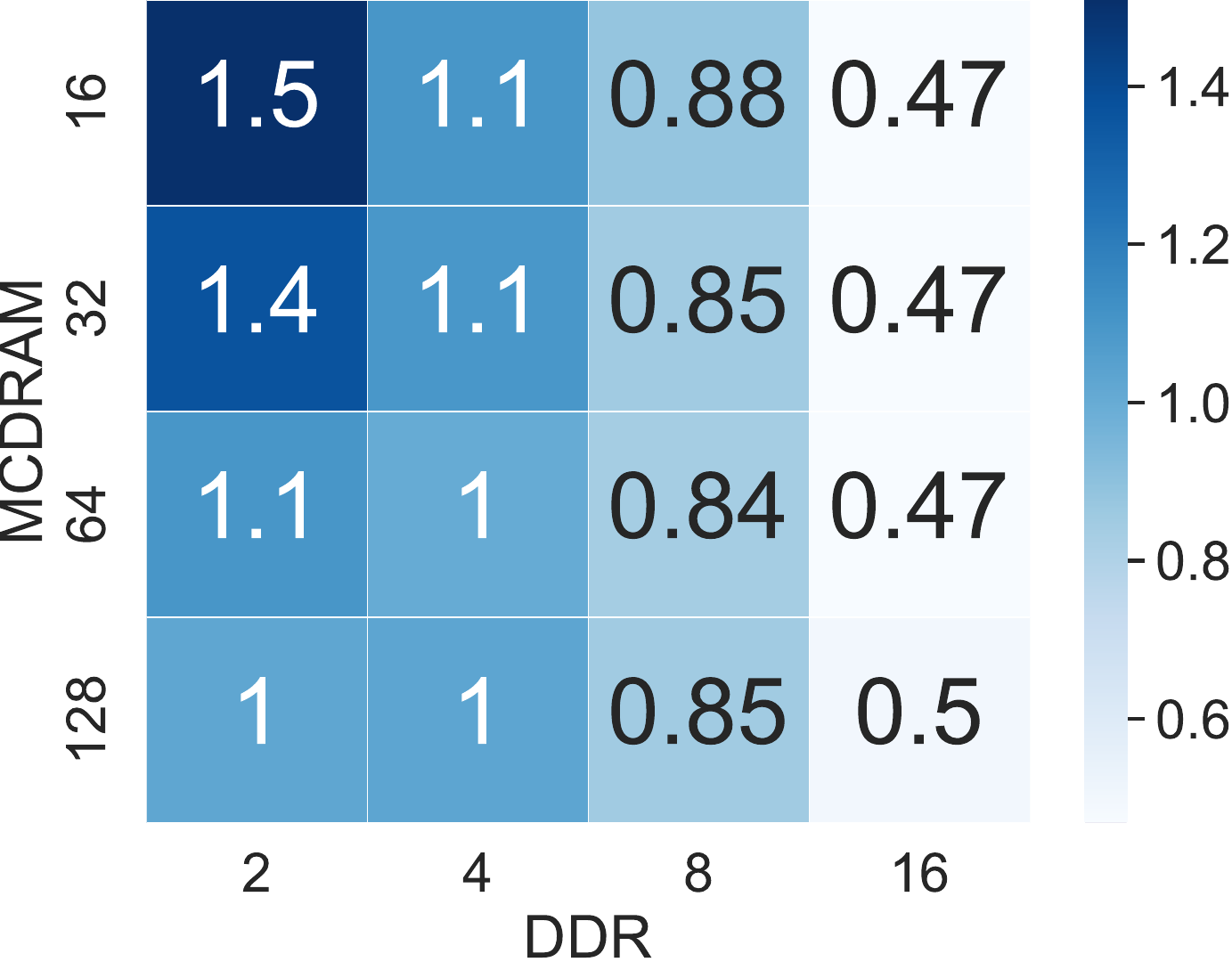}
\caption{[15-16]}\label{fig:31GB_time}
\end{subfigure}
\caption{(a) \& (b) Execution time [s] of STREAM Triad with data distribution [X-Y], where X = GBs placed in MCDRAM and Y = GBs placed in DRAM}
\label{fig:pc_heatmaps}
\end{minipage}
\begin{minipage}{0.49\linewidth}
\centering
    \includegraphics[width=0.9\columnwidth]{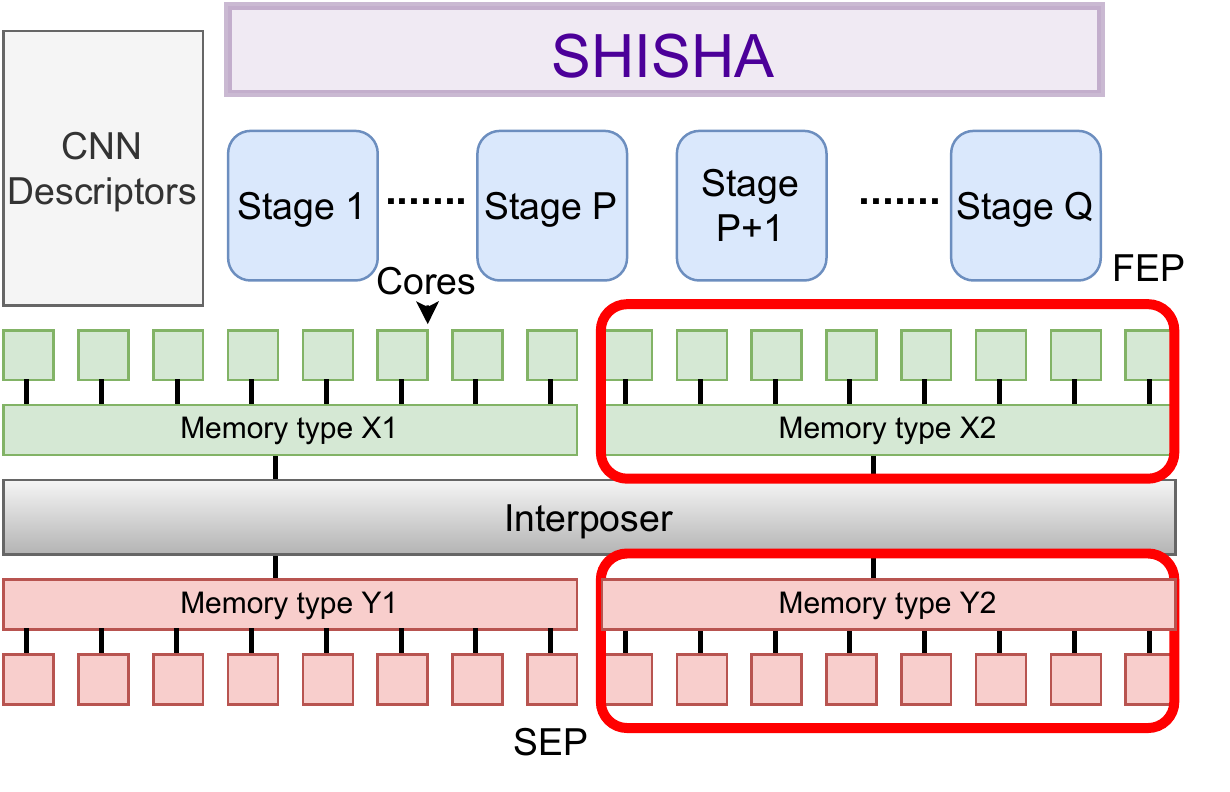}
    \caption{System targeted in this paper. Memory type X and Y represent different memory bandwidths.}
    \label{fig:shisha_system}
\end{minipage}
\vspace{-0.5cm}
\end{figure}

\vspace{-.3cm}
In a computing platform with different types of memories, the assignment of workload and data objects becomes crucial for better performance. 
To investigate the impact of different thread and data assignment strategies, we tested the STREAM Triad benchmark~\cite{mccalpin1995stream} with two data sizes, 19 GB and 31 GB on Intel's Knights Landing (KNL)~\cite{sodani2015knights}. 
KNL has two types of memories, 16 GB of high bandwidth memory(HBM), also called MCDRAM, and 90 GB of DDR4 DRAM. The bandwidth of HBM is $~4\times$ higher than that of DRAM~\cite{salehian2017evaluation}. This suggests that most of the application data should be placed in HBM. 
It also means that HBM should be able to handle more parallelism until the bandwidth is saturated.
For each data size, 15 GB of data are placed in MCDRAM and the remainder of the data are placed in DRAM. 
In Figure \ref{fig:comparisons}, we show three cases, namely, 1) when all data are placed in DRAM (DDR only),   
2) when MCDRAM is used as a cache (cache mode), and 
3) when data is distributed across the two memories.
As can be seen, with a sensible thread assignment, the case 3 yields the best performance. 
This shows that a clever data partitioning and thread assignment are key to achieve high performance in the presence of memory heterogeneity.
Further analyzing case 3, 
Figure \ref{fig:pc_heatmaps} shows the heatmap of the execution time of STREAM Triad with different thread assignments to MCDRAM $[16,32,64,128]$ and DRAM $[2,4,8,16]$. 
The optimal number of threads is determined by a) the memory bandwidth of each memory type, b) the additional bandwidth consumed by each extra thread, and c) the amount of data to be processed. 
Results from the experiment show that for each data partitioning between HBM and DRAM there is a different optimal thread partitioning. 
An important observation from Figures \ref{fig:pc_heatmaps} is that better performance can be achieved by assigning fewer number of threads per memory type, rather than opting for assigning maximum number of threads. 

\noindent
\textbf{Problem definition and general approach of the solution:}\\ 
This work considers a computing platform which is composed of 
a set of \rd{nodes} 
consisting of high performance cores attached to a high-bandwidth memory (referred to as Fast Execution Place -- FEP) and clusters of relatively slower cores attached to a low-bandwidth memory (referred to as Slow Execution Place -- SEP). This MCM based scenario is expected for chiplet architectures with heterogeneous integration and is shown in Figure \ref{fig:shisha_system}. Our goal is to run throughput maximizing CNN inference pipelines on such an architecture. 


%% file: CP38_background.tex
\vspace{-.5cm}
\section{Background and related work}\label{sec:background}\vspace{-.3cm}
There are various schemes for parallelizing CNNs.
In data parallelism the work of a minibatch (a set of inputs) is partitioned among multiple computational resources. 
In model parallelism the work is divided according to neurons in each layer which corresponds to the tensor dimensions in each layer. 
In layer pipelining~\cite{ben2019demystifying} the work is partitioned by distributing network layers among computational resources. Model parallelism within the layer is combined with layer pipelining by arranging computational resources into multiple teams of workers. This hybrid parallelism has following benefits:
1) there is no need to replicate weight and input tensors on all devices, 
2) the communication volume and points are reduced, and 
3) the weights can remain cached, thus decreasing memory round-trips. 
In the rest of the paper we will refer to CNN pipelines in which network layers are grouped into pipeline stages. Each pipeline stage is assigned a unique set of computational resources, referred to as EPs.

Finding out the right schedule and mapping of CNN pipelines on mentioned architectures is a design space exploration problem, where we are interested in the configuration that achieves the highest throughput. The configuration consists of the number of pipeline stages, CNN layers per pipeline stage and a mapping of pipeline stages to EPs. In the literature, various 
meta heuristic and machine learning algorithms have been used such as Simulated Annealing~\cite{zheng2020ansor}, evolutionary algorithms~\cite{soomro2021online, adams2019learning}, reinforcement learning~\cite{ahn2020chameleon,oren2021solo} and deep neural network techniques~\cite{anderson2021efficient}.
The design space under consideration is large and complex, requiring tens of thousands of trials in order to reach a near optimum with current search schemes.\rd{Exploring in such a complex space is NP-hard}. 

%% file: CP38_relatedWork.tex
Parallel pipelines for CNN training have been applied in practice~\cite{fan2021dapple, narayanan2019pipedream, narayanan2021memory,huang2019gpipe}. Recently, Chimera~\cite{li2021chimera} generates a schedule for bi-directional pipelines by using complex cost models that represent the execution time of one network pass and calculate the depth and parallelism per pipeline stage. 
In Halide,~\cite{adams2019learning} the pipeline scheduling approach uses a cost model that considers 66 platform and application specific features. For the cost model, 26 out of 66 feature values are predicted by a neural network trained on random representative programs. According to the specifications, one training point takes at most 320 minutes to train the neural network using different schedule configurations. To predict a schedule for Halide pipelines of a single CNN layer, the scheduler considers 10k configurations. In comparison, we show that for a large YOLOv3 network of 52 layer, \shisha\ 
considers only 18 configurations.

%% file: CP38_methodology.tex
\vspace{-.4cm}
\section{\shisha\ exploration approach}\label{sec:shisha}\vspace{-.2cm}
A pipeline configuration consists of two components:  
1) the number of CNN layers assigned to each pipeline stage, and 
2) the assignment of each pipeline stage to an EP.
An EP can be a single or multiple cores attached to a memory module. Therefore, we classify the EPs according to the type of memory. For example, in Figure \ref{fig:shisha_system} EPs are colored in green or red. We use this classification in \shisha\ to provide hints about the characteristics of the computing platforms with heterogeneous modules. 

\shisha\ is a two-step approach. 
The first step is the ``seed generation'', in which we use a simplified cost-model to come up with an initial solution.
This initial solution is used in the second step, ``online tuning'' for faster convergence.
\vspace{-.7cm}
\subsection{Seed generation}\label{sec:seed}\vspace{-.3cm}
The goal of the seed generation is to determine a sensible starting configuration using only static information.

Firstly, Equation~\ref{eq1:conv_eq} is used to calculate the weights of the CNN layers \cite{AUGUR, pipeit, CSDF, graphi}. For each layer, $H,W,C$ denote the height, width and depth of the input tensor. $R,S$ represent the height and width of the underlying convolutional kernel and $K$ is the number of filters of the convolutional kernel. Note that, we are considering conventional CNNs in this paper, other type of layers can be incorporated in the context of this work by replacing Equation~\ref{eq1:conv_eq} with a model for the estimation of computational intensity of the layers.  
\vspace{-0.3cm}
\begin{equation} \label{eq1:conv_eq}
\vspace{-0.3cm}
    W = H \times W \times C \times R \times S \times K
\end{equation}
Secondly, we capture the heterogeneity of the system to support the seed generation. This is used to guide the mapping of pipeline stages to EPs together with the total weight of each pipeline stage. We rank the EPs in a decreasing order of performance, for example, from Figure \ref{fig:shisha_system} green EPs have rank 1 (FEP) and red ones have rank 2 (SEP). This is a hint to \shisha\ to balance the workload considering static knowledge about the heterogeneity of the system.

\begin{figure*}[ttt!]

\begin{minipage}[t]{.48\textwidth}
\begin{algorithm}[H]
\begin{algorithmic}[1]
        \REQUIRE $W_l, H_e, N, L, C$
        \STATE $seed[N]$
        \STATE $E[N]$
        \FOR{$passes$ in [0..$|L-N|$]}
        \STATE $min_w \leftarrow min(W_l)$
        \STATE $n \leftarrow min(min_w-1, min_w+1)$
        \STATE $W_l \leftarrow merge(min_w, n)$
        \STATE $seed \leftarrow merge\_layers(min_w, n)$
        \ENDFOR
        \STATE $Rank \leftarrow rank(seed, W_l, C)$
        \FOR{$i$ in $[0...N]$}
        \STATE $E[Rank_i] \leftarrow assign(Rank_i,H_{ei})$
        \ENDFOR
        \RETURN $seed, E$
        \vspace{1.16cm}
\end{algorithmic}
    \caption{Seed Generation}
    \label{alg:seed}
\end{algorithm}
\end{minipage}
 \begin{minipage}[t]{.48\textwidth}
\begin{algorithm}[H]
\begin{algorithmic}[1]
        \REQUIRE $seed, E, H_e, \alpha$
        \STATE $conf \leftarrow seed$
        \STATE $throughput = execute(conf)$
        \STATE $\gamma \leftarrow 0$
        \WHILE{$\gamma$ \textless $\alpha $}
            \STATE $stage \leftarrow slowest\_stage(conf)$
            \STATE $t\_stage \leftarrow nearestFEP(E)$
            \STATE $conf \leftarrow move(conf, t\_stage)$
            \STATE $Tp = execute(conf)$
            \IF{$Tp \leq throughput$}
                \STATE $\gamma++$
            \ELSE
                \STATE $\gamma \leftarrow 0$
                \STATE $throughput \leftarrow Tp$
            \ENDIF
            \ENDWHILE
        \RETURN $conf$
    \end{algorithmic}
    \caption{Online Tuning}
    \label{alg:tuning}
\end{algorithm}
 \end{minipage}

\vspace{-0.7cm}
\end{figure*}

The seed generation process is described in Algorithm~\ref{alg:seed}. $W_l = [w_{l1}, w_{l2}, .... w_{lL}]$ is the weight list, where a layer weight $w_{li}$ is calculated using Equation~\ref{eq1:conv_eq}. $H_e = [e_1, e_2,... e_N]$ is a list of EPs sorted in descending order w.r.t. performance. For example, for Figure \ref{fig:shisha_system} $H_e = [G_1,G_2, ..G_p,  R_1, R_2, ..R_q]$ represents the $p$ EPs that belong to memory types X (green) and $q$ Y (red) EPs. $L$ is the total number of layers in a given CNN. $N$ is the total number of pipeline stages in final pipeline ($N\leq L$) and $C$ is assignment choice which is discussed later in this section.
The output of Algorithm \ref{alg:seed} is a pipeline configuration $Seed = [PS_1, PS_2, ... PS_N]$, where $PS_i$ represents the number of CNN layers assigned to $i_{th}$ pipeline stage. Output $E=[e_1, e_2, ...e_N]$ is a list of EPs from $H_e$ and the corresponding assignment to pipeline stages. Algorithm \ref{alg:seed} comprises two phases. In phase 1 (Lines from 3-8) we generate pipeline stages by combining CNN layers. The goal of this phase is to merge layers into groups in order to balance out the cumulative weight of groups. These groups eventually become pipeline stages. The idea is to look for the layer with lowest weight (Line 4) and merge it with the immediate neighbour with the smallest weight (Line 5,6). Typically, the weight distribution in CNN layers does not follow any order, i.e. a light weight layer can be found between two layers with heavy weights. 
The second phase of Algorithm \ref{alg:seed} (Lines 9-11) assigns the pipeline stages output by phase 1 to EPs. In principle, heavy pipeline stages should be assigned to high performance EPs, however, the assignment is not trivial in practice and requires to examine the impact of a few heuristics. Eventually, this will help in balancing execution time per pipeline stage, thus achieving a balanced pipeline. 

\textbf{Stage-to-EP assignment heuristics}\label{sec:assignment_schemes}:  
Once CNN layers are grouped into pipeline stages, we then 
assign an EP to each pipeline stage. Since we have information about performance heterogeneity among EPs, we can make different choices, such as;  
    1) Rank pipeline stages w.r.t. number of \emph{layers} assigned to each pipeline stage ($Rank_{l}$). While merging layers into stages, it is sometimes inevitable to have pipeline stages which are heavy in terms of aggregated weight with many light weight layers as opposed to a pipeline stage with one heavy layer. The highest rank corresponds to the pipeline stage with highest number of layers. We assign higher ranks to SEPs. This facilitates the online tuning phase later to greedily move the layers among pipeline stages to reach a solution. 
    2) Rank pipeline stages w.r.t. aggregated \emph{weight} of each pipeline stage ($Rank_{w}$) Here, we assign the pipeline stages with heavy weights to fast EPs to balance the load.
Line 9 controls this choice in Algorithm \ref{alg:seed}. 
%
\vspace{-.4cm}
\subsection{Online tuning} \label{sec:tuning}\vspace{-.2cm}
For the exploration phase,  
we strive to reduce the exploration time so that it is still practical to carry out an online exploration without causing a significant overhead on execution time. 
This is particularly challenging given the size of the multidimensional pipeline configuration space, which often includes an overwhelming majority of slow configurations.
We avoid visiting such configurations by starting from the seed configuration and incrementally adjusting load distribution by moving layers from one pipeline stage to an adjacent lighter stage. 
In Algorithm \ref{alg:tuning}, we describe the auto-tuning scheme of \shisha\ . The required input is a pipeline configuration generated as a seed. A list of EPs $E$ which represent a mapping of pipeline stages to the computing platform. 
The $\alpha$ parameter controls how many configurations are attempted after a configuration that outperforms the seed and recently found solution has been detected. 
The rationale behind Algorithm \ref{alg:tuning} is to gradually reduce the load of the slowest pipeline stage in order to improve the overall throughput of the pipeline. 
Hence, \shisha\ finds the slowest stage (Line 5) and remaps one layer at a time to the nearest faster EPs (Line 6-7). \rd{The layer could be popped from front or back end of the stage depending on the location of new EP}. 
Once a better configuration is found than any previous one, we try $\alpha$ more times to search for a better configuration. 
In Line 6 we balance the workload by moving layers to a nearest fast EP ($nFEP$) in pipeline \rd{i.e a closer stage which is running on an FEP}. 
However, this is not the only choice that can be made. 
The nearest lightest fast EP ($nlFEP$) is also a good target to move layers as well. Therefore we keep both options open for the user to select. The complexity of \shisha\ is negligible therefore it does not cause much work to test different choices for a given CNN and computing platform.

%% file: CP38_experimental_setup.tex
\vspace{-.4cm}
\section{Experimental setup}\label{sec:ExpSetup}\vspace{-.2cm}
\shisha\ targets systems that are heterogeneous in core performance and memory bandwidth. As discussed in Section \ref{sec:motivation}, the system under consideration consists of different types of cores attached to different memory modules. Chiplets such as Nvidia's Simba~\cite{shao2019simba} and Intel Meteor Lake~\cite{intel_2022} resemble such types of architectures.
We used the gem5 simulator~\cite{binkert2011gem5} to simulate heterogeneous cores and memory bandwidth. The simulator provides flexibility in modeling different architectures.
To simulate different core performances, we used ARM's bigLittle cores~\cite{greenhalgh2011big} models in gem5 and to simulate different memory types, we tried different memory bandwidth values using a simple memory model connected to core cluster in gem5.
Inter-EP latency is set to $20ns$~\cite{shao2019simba}. However, the execution time of pipeline stages is orders of magnitude higher than inter-EP latency, thus it does not impact the performance of pipeline.

A GEMM-based implementation~\cite{darknet13} consists of two operators; 1) Im2Col and 2) GEneralized Matrix Multiplication (GEMM). We include both operators to simulate execution time for CNN layers of ResNet50, YOLOv3 and AlexNet.

%% file: CP38_evaluation.tex
\vspace{-.5cm}
\section{Evaluation}\label{sec:evaluation}\vspace{-.3cm}
As highlighted previously, \shisha\ includes a seed generation component and an online tuning heuristic.
In this section, we evaluate the quality of the seed and the final solution generated by \shisha\ and analyze the convergence of the online auto-tuning phase. 

Pipe-Search~\cite{soomro2021online} is an online approach that uses a database of pipeline configurations sorted w.r.t. the distribution of workload among pipeline stages. 
It tests pipeline configurations of various depths and converges to a solution when no better solution is found by a time limit set by the user. 
This approach incurs a high overhead when generating the database of pipeline configurations which also limits its scalability. 
We compare \shisha{}'s auto-tuning module with a 
set of exploration algorithms commonly used in literature, such as Hill Climbing (HC) \rd{with proximity equal to the number of layers in the network}, Simulated Annealing (SA) \rd{with cooling factor values ranging from $0.9^{-5} - 0.01$}, Random walk (RW) and in selected cases, Exhaustive Search (ES). For a fair comparison we test SA and HC with seeds produced by \shisha\ referred to as SA$_s$ and HC$_s$. \rd{For randomized algorithms, we run 200 times and picked the solution which is closer to near optimal value}

We use three CNNs in our experiments. ResNet50~\cite{he2016deep} and YOLOv3~\cite{redmon2018yolov3} are widely used image classification CNNs. There are $50$ compute intensive layers in ResNet50 and $52$ compute intensive layers in YOLOv3. The generation of sorted configurations, as required by Pipe-Search and ES, incurs an impractical time overhead when running ResNet50 and YOLOv3 for more than 4-stage pipelines. 
Therefore, we extend our benchmark set with a synthetic network (SynthNet) consisting of $18$ convolutional layers. These layers are taken from the AlexNet architecture
as AlexNet has only five convolutional layers and our testing platform consists of 8 EPs.
This is to analyze CNNs that can be run on a higher number of EPs (i.e. $EP>8$) and have a compute complexity matching widely used CNNs. 
\vspace{-.5cm}
\subsection{Comparison of \shisha\ with exploration algorithms}
\begin{figure}[!tb]

\centering
\begin{minipage}[t]{.45\textwidth}
    \centering
     \includegraphics[width=0.95\linewidth]{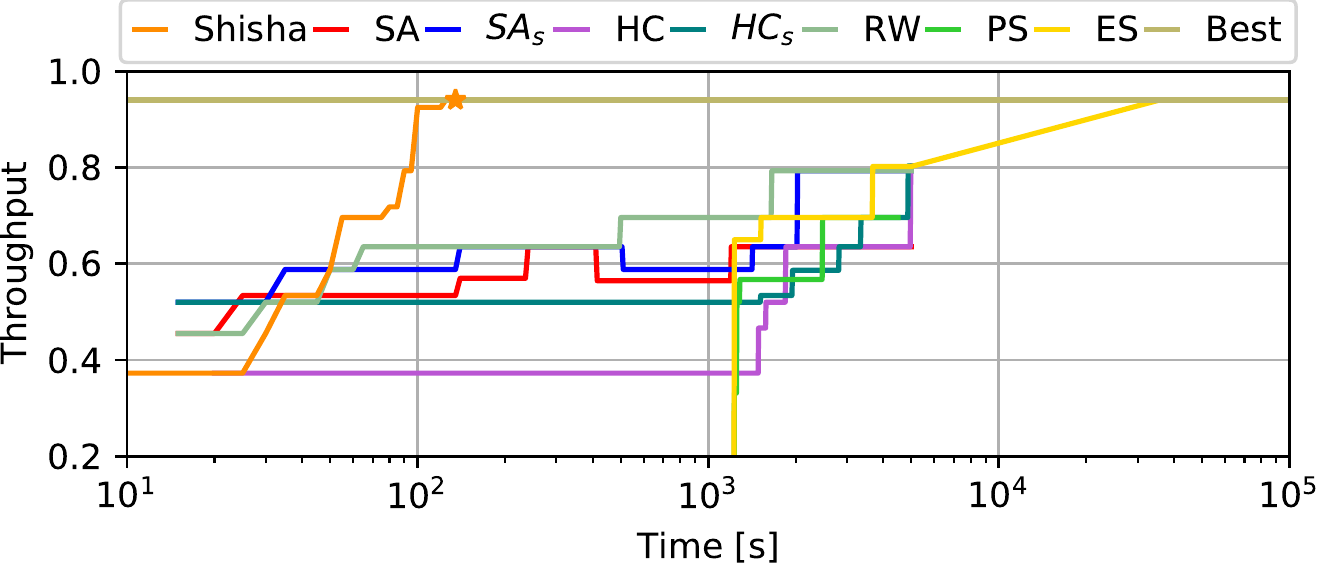}
    \caption{Convergence of exploration algorithms for SynthNet on 8 EPs. Xaxis is time in log scale}
    \label{fig:solution_comparison}
\end{minipage}%
\hspace{0.3cm}
\begin{minipage}[t]{.45\textwidth}
  \centering
     \includegraphics[width=0.95\linewidth]{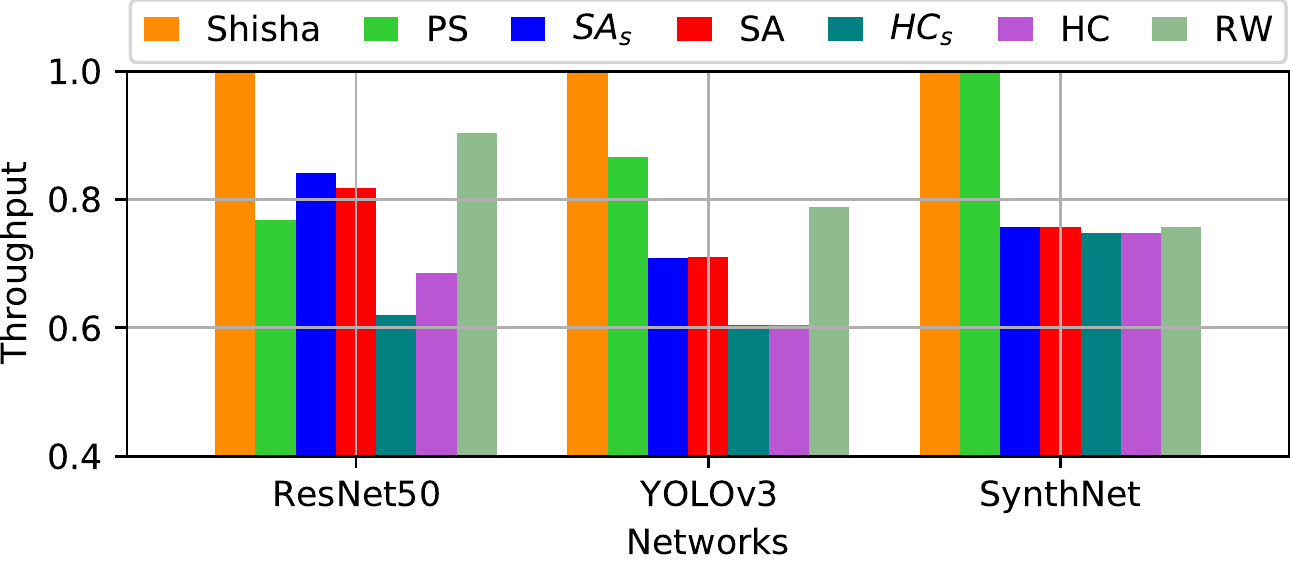}
    \caption{Throughput of search schemes normalized to ES}
    \label{fig:exhaustive}
\end{minipage}
\vspace{-0.9cm}
\end{figure}

\vspace{-.3cm}
Figure \ref{fig:solution_comparison} shows the convergence behavior of all exploration algorithms. The solution found by \shisha\ is equal to the best solution found by ES. For a fair comparison we run SA and HC using the same seed ($SA_s$, $HC_s$) generated by \shisha\ as a starting configuration. 
HC tries configurations in close proximity; both versions of $HC$ and $SA$ managed to find a better solution ($throughput$ $=$ $0.80$) compared to the best solution ($throughput$ $=$ $0.94$).
However, the time of convergence of representative exploration approaches is high, this is because of using many configurations out of which some are very slow. ES and PS, on the other hand, incur the overhead of generating a database of configuration. As shown in Figure \ref{fig:solution_comparison}, it took $~1200s$, after that ES and PS started exploring.  
\shisha\ explores $0.12\%$ of the total design space as compared to Pipe-search which explores $2.03\%$ of the design space.
\rd{this is because \shisha\ attempts configurations which leads towards the solution faster}
On average, the convergence time is improved by $\sim35\times$ in \shisha\ compared to other search algorithms.
In our approach, the stopping condition is controlled by $\alpha$ as mentioned in Section \ref{sec:tuning}. We used $\alpha=10$ in our experiments.


\vspace{-.5cm}
\subsection{Analysis of optimality}\vspace{-.3cm}
To quantify the confidence on \shisha\ solutions, we compared against ES using larger CNNs. In this experiment we configured a system of four EPs as it takes a lot of time for ResNet50 and YOLOv3 to run ES for higher number of EPs.
Figure \ref{fig:exhaustive} shows the throughput ($=1/(ExecutionTime\,of\,slowest\,stage)$) of the solution found by \shisha\ and other algorithms normalized to best solution found by ES. In case of ResNet50 and YOLOv3, \shisha\ found the best solution by exploring $~0.1\%$ of the design space. In case of SynthNet, \shisha\ explored $2.5\%$ of the design space to find the best solution. This is due to the fact that design space of SynthNet (18 layers) is smaller than ResNet50 (50 layers) and \shisha\ on average tries $25-35$ exploration points with $\alpha = 10$. 
\vspace{-.5cm}
\subsection{Importance of seed in the auto-tuning phase of Shisha}
\begin{figure}[!tb]
\begin{minipage}[t]{.48\linewidth}
\centering
\begin{subfigure}[b]{.49\linewidth}
\includegraphics[width=\linewidth]{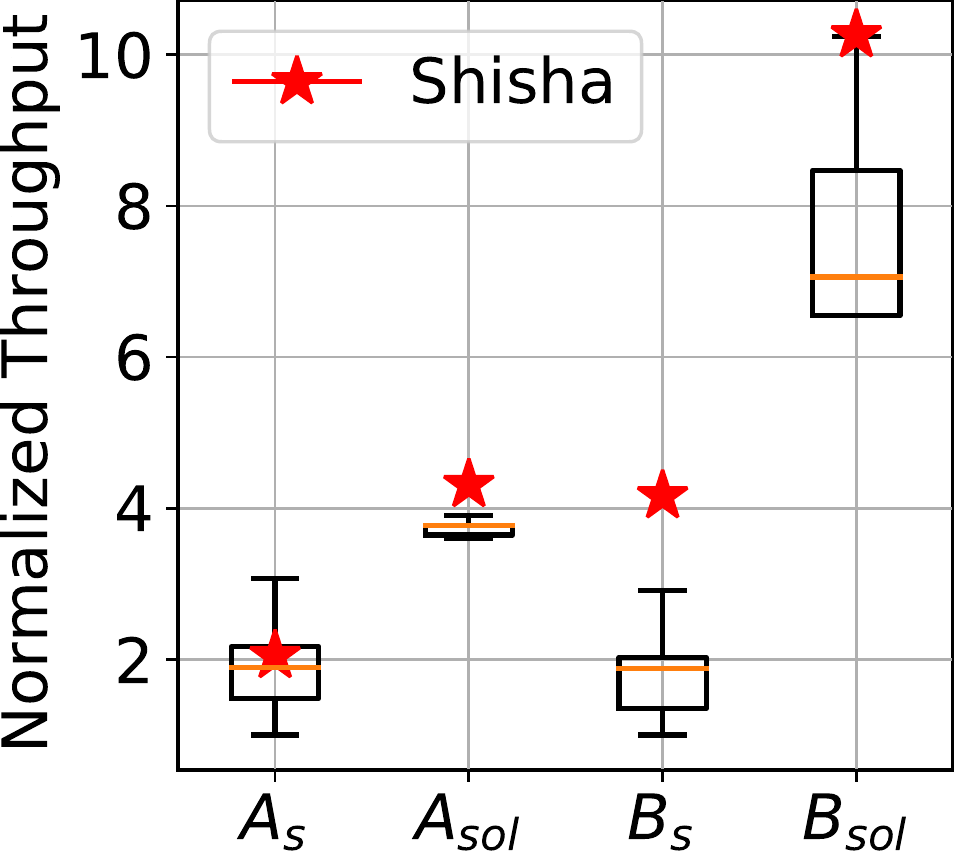}
\caption{Throughput}\label{fig:seed_sol_tp}
\end{subfigure}
\centering
\begin{subfigure}[b]{.49\linewidth}
\includegraphics[width=\linewidth]{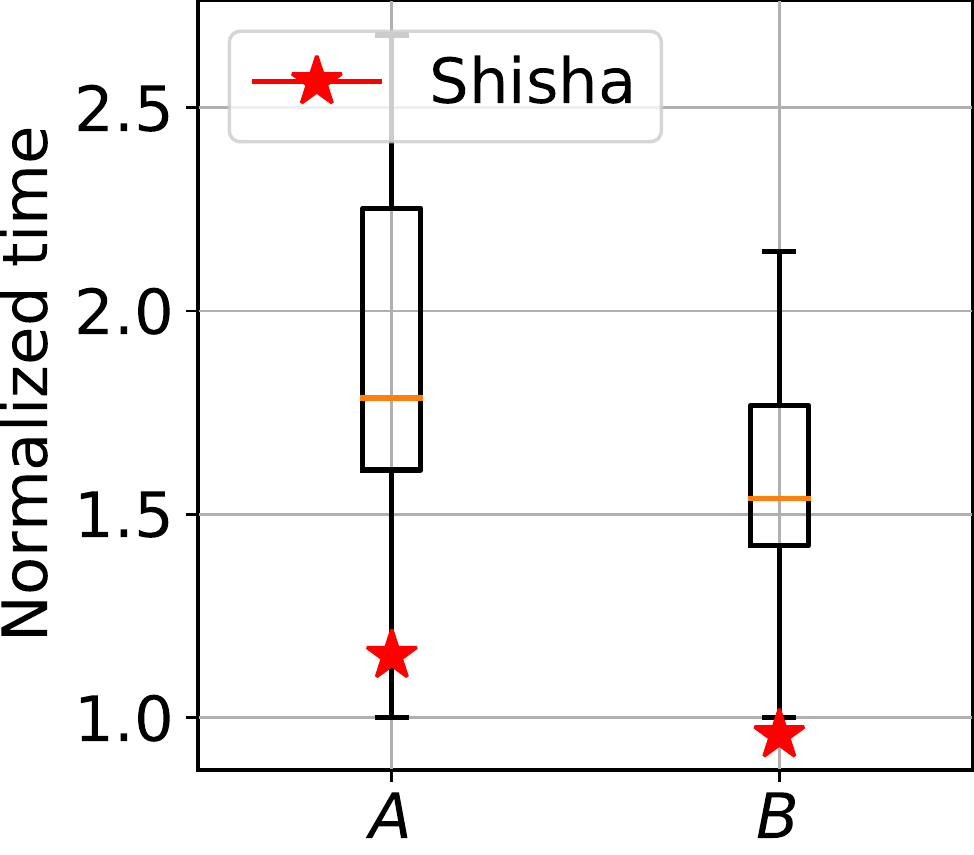}
\caption{Convergence}\label{fig:seed_sol_et}
\end{subfigure}
\caption{Comparison of \shisha\ seed against a set of 100 random seeds. $s$ = seed, $sol$ = solution, $A$ = YOLOv3 and $B$ = ResNet50.}
\label{fig:seedVsSol}
\end{minipage}
\hspace{0.3cm}
\begin{minipage}[t]{.49\linewidth}
\centering
\begin{subfigure}[b]{.49\linewidth}
\includegraphics[width=\linewidth]{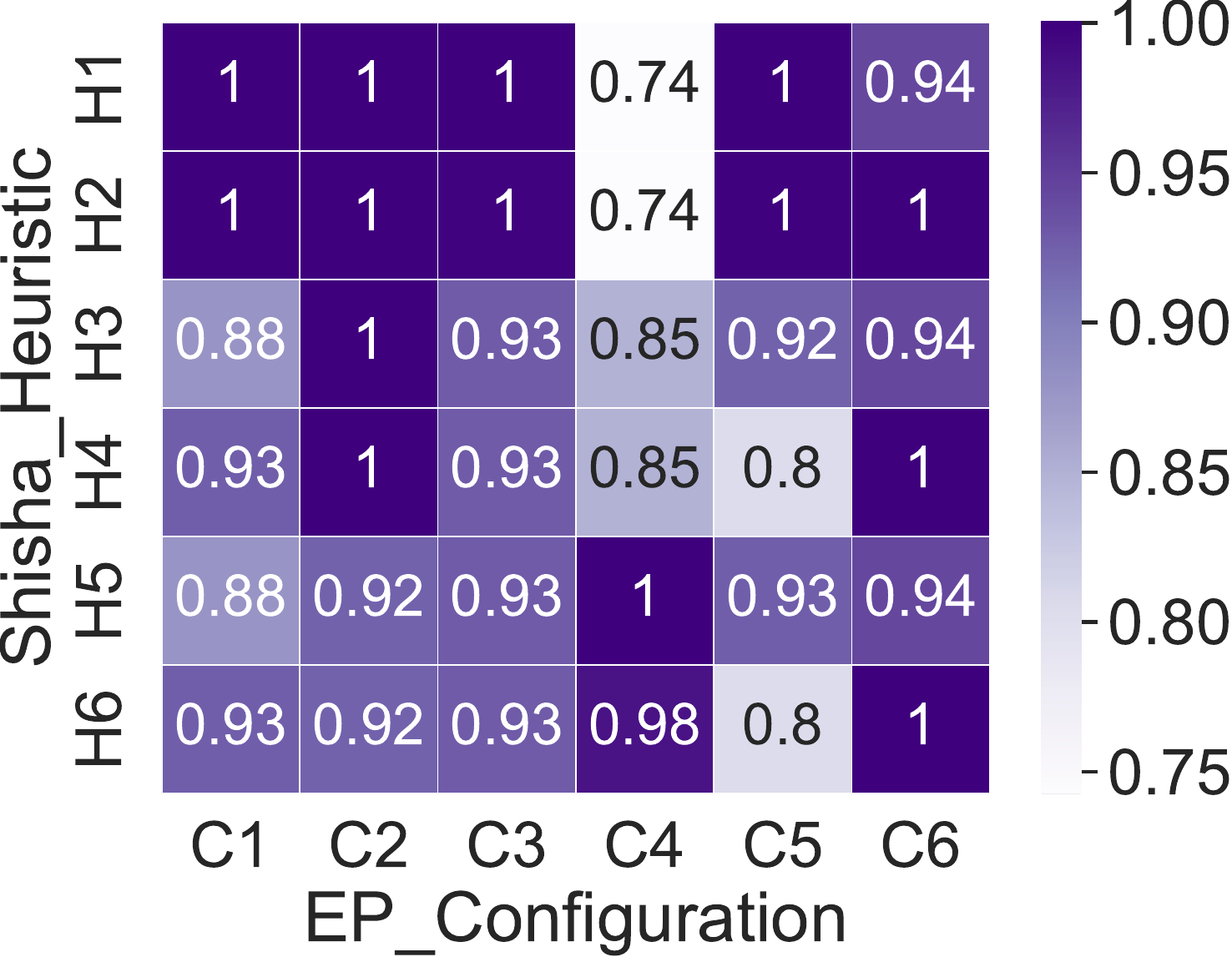}
\caption{ResNet50}\label{fig:rn50}
\end{subfigure}
\centering
\begin{subfigure}[b]{.49\linewidth}
\includegraphics[width=\linewidth]{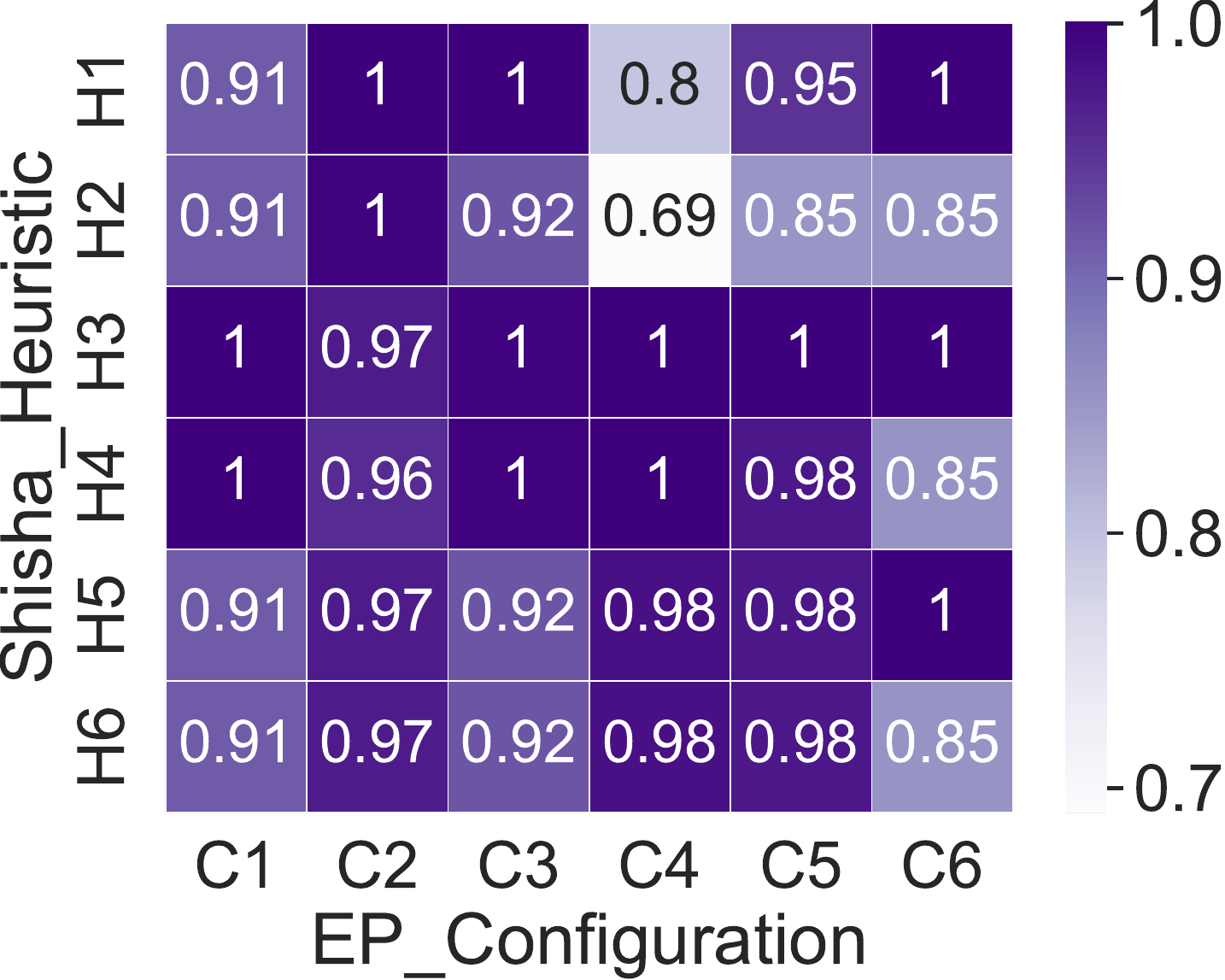}
\caption{YOLOv3}\label{fig:yolo}
\end{subfigure}
\caption{Throughput using different heuristics~\ref{tab:versions} and configurations of EPs~\ref{tab:ec}}
\label{fig:Shisha_trajectory}

\end{minipage}
\vspace{-.7cm}
\end{figure}

\vspace{-.3cm}
The seed generated by \shisha\ contains the mapping of pipeline stages to EPs.
Figure \ref{fig:seedVsSol} represents the throughput and convergence time of \shisha\ when initiated with the seed generated by Algorithm \ref{alg:seed}, represented as \shisha\ mark(red), compared to a set of 100 random seeds and solutions obtained with random seeds. 
In case of ResNet50, the solution quality in both cases is similar but convergence time is increased by $35\%$ when started with a random seed. In case of YOLOv3, the throughput of the solution found using \shisha\ seed is $16\%$ better and the convergence time is always better than a solution found using a set of 100 random seeds.

\vspace{-.5cm}
\subsection{Assignment and balancing schemes in \shisha\/}\vspace{-.2cm}
Section \ref{sec:assignment_schemes} and \ref{sec:tuning} discuss various choices that \shisha\ makes while assigning EPs and balancing workload among pipeline stages. We investigate the impact of each of these choices, with results shown in Figure \ref{fig:Shisha_trajectory}. Table \ref{tab:versions} lists the heuristics to be configured in \shisha\/. Assignment of EPs in H5 and H6 is random, in order to study the impact on convergence when no heuristic is used. Table \ref{tab:ec} lists various configurations of the computing platform used to run this sensitivity analysis. The balancing scheme \textit{lightest FEP} is effective in all cases as \shisha\ tries to move workload to an FEP which takes least time to execute assigned pipeline stage. This helps in balancing the pipeline as well as maximizing the throughput of the pipeline. In $80\%$ of the cases, H1 and H3 yield better results. 
We investigated the convergence time of both schemes in order to determine the effectiveness of H1 and H3. Figure \ref{fig:h1_h3} Shows that the convergence time of H3 is less than H1 in $90\%$ of the cases. This is due to the fact that in H3 assignment is done w.r.t. weights which means the configurations tested during exploration take reasonably less time than in H1. We recommend to use H3 because it converges faster and yields a near optimal solution.
\begin{table}
\vspace{-0.7cm}
\begin{minipage}[t]{.6\textwidth}
\centering
 \scalebox{0.7}{
 \begin{tabular}{@{}c|c|c@{}}
\toprule
\textbf{Heuristic \#} & \textbf{Assignment of EPs} & \textbf{Balancing}      \\ \midrule
H1                & $Rank_{l}$                           & $nlFEP$ \\
H2                & $Rank_{l}$                           & $nFEP$  \\
H3                & $Rank_{w}$                          & $nlFEP$ \\
H4                & $Rank_{w}$                          & $nFEP$  \\
H5                & random                                  & $nlFEP$ \\
H6                & random                                  & $nFEP$  \\ \bottomrule
\end{tabular}
}
\caption{Heuristics of \shisha}
\label{tab:versions} 
\end{minipage}%
\begin{minipage}[t]{.4\textwidth}
\centering
 \scalebox{0.7}{
\begin{tabular}{@{}c|c|c@{}}
\toprule
\textbf{Conf.} & \textbf{FEPs} & \textbf{SEPs}      \\ \midrule
C1                                 & 1 8-core                          & 1 8-core                          \\
C2                                 & 2 8-core                          & 2 8-core                          \\
C3                                 & 4 4-core                          & 2 8-core                          \\
C4                                 & 2 8-core                          & 4 4-core                          \\
C5                                 & 4 4-core                          & 4 4-core                          \\ 
C6                                &  8 4-core                        &   NULL                    \\ 
\bottomrule
\end{tabular}
}
\caption{EPs}
\label{tab:ec}
\end{minipage}

\end{table}
\vspace{-1.2cm}
\vspace{-.5cm}
\subsection{Sensitivity analysis of $\alpha$}\label{sec:interfernce}\vspace{-.3cm}

\begin{figure}
\begin{minipage}{0.4\linewidth}
    \centering
     \includegraphics[width=\columnwidth,keepaspectratio, height=2.5cm]{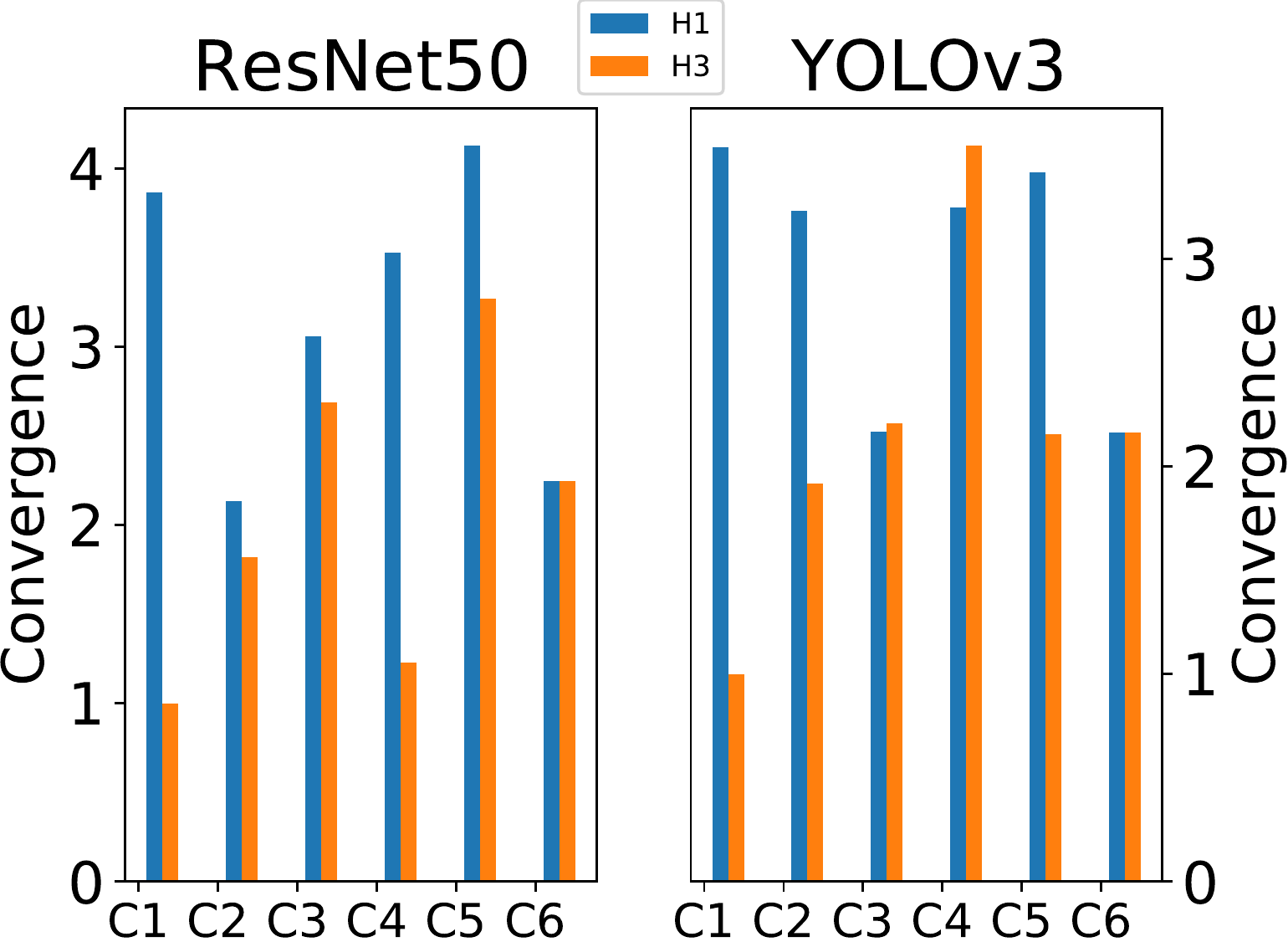}
    \captionof{figure}{Convergence time normalized to minimum value in each group for H1 and H3.}
    \label{fig:h1_h3}
\end{minipage}
\hspace{0.04cm}
\begin{minipage}[]{0.58\textwidth}

\centering
     \includegraphics[width=\columnwidth]{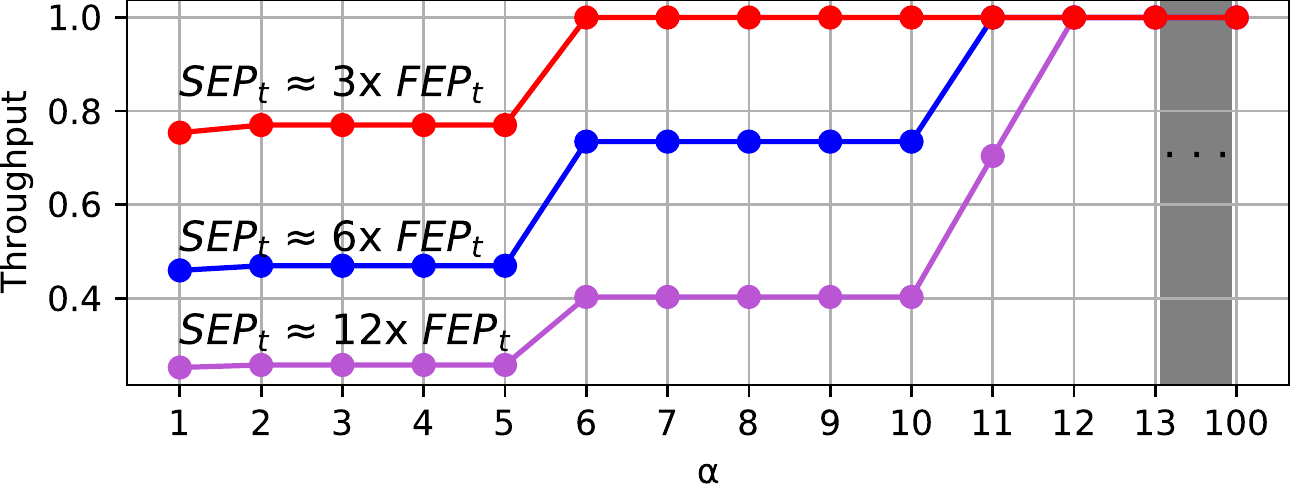}
    \caption{Impact of $\alpha$ on the quality of solution in presence of heterogeneity}
    \label{fig:alpha}
\end{minipage}

\vspace{-0.7cm}
\end{figure}

The extent of exploration of \shisha\ is controlled by $\alpha$. The value of $\alpha$ should be chosen such that it allows tuning according to the performance heterogeneity among FEPs and SEPs while keeping a sensible convergence time. A higher value of $\alpha$ also means a longer tuning phase. Figure~\ref{fig:alpha} shows the quality of solution (normalized to throughput obtained when $\alpha = 100$) for the YOLOv3 pipeline tested on three platform configurations with the SEPs $[3\times,\, 6\times,\,12\times]$ slower than the FEPs. In our experiments, the performance difference between ARM's Big and Little cores is three folds on average, which is the first case in the figure. 
It is shown that with the higher heterogeneity between EPs, higher $\alpha$ yields a better solution. 
We use the same starting seed for the same CNN in all cases, therefore, for lower values of $\alpha$, throughput behavior is similar, irrespective to the performance difference between EPs, but in the case of a higher performance difference, throughput is improved with a higher value of $\alpha$.

%% file: CP38_conclusion.tex
\vspace{-.5cm}
\section{Conclusion}\label{sec:conclusion}\vspace{-.3cm}
In this work we demonstrate a fast approach to scheduling CNN pipelines on heterogeneous computing platforms consisting of fast and slow cores. The proposed approach is generic and can be used on platforms featuring GPUs or FPGAs, in addition to asymmetric multicores and chiplets. We utilize compile time information in combination with a brief and guided online search for auto-tuning the CNN layers into parallel pipelines. Our experimental evaluation shows that the solution found by \shisha{} is as good as one produced by an exhaustive search of the design space. The results also show that \shisha{} scales well with larger networks and computing platforms.  
In future work, we will look at more generic tensor expressions~\cite{rink_array19} and the effect on seed parameters of high-level algebraic transformations~\cite{rink_gpce18}. 

\section*{Acknowledgment}\label{sec:acknowledgment}
This work has received funding 
from the EU Horizon 2020 Programme under grant agreement No 957269 (EVEREST), 
from the AI competence center ScaDS.AI Dresden/Leipzig (01IS18026A-D),
PRIDE from Swedish Foundation for Strategic Research with reference number CHI19-0048 and
eProcessor from the European High-Performance Computing Joint Undertaking (JU) under grant agreement No 956702.
Some of the computations were enabled by resources provided by the Swedish National Infrastructure for Computing (SNIC) at Chalmers Centre for Computational Science and Engineering (C3SE) partially funded by Swedish Research Council https://www.vr.se/ under grant agreement No 2018-05973.